%% file: coleman3.tex
\documentclass[twoside]{ahp}
\usepackage{coleman_h}
\usepackage{amsmath}
\usepackage{graphicx}
\usepackage{epsfig}
\newlength{\upit}\upit=0.1truein

\newcommand{\ltappr}{{{\lower4pt\hbox{$<$} } \atop \widetilde{ \ \ \ }}}
\newlength{\bxwidth}\bxwidth=1.5 truein

\begin{document}

\newcommand{\dg}{^{\dagger }}
\newcommand{\si}{\sigma}
\newcommand{\rarrow}{\rightarrow}
\def\fig#1#2{\includegraphics[height=#1]{#2}}
\def\figx#1#2{\includegraphics[width=#1]{#2}}
\newlength{\figwidth}
\newlength{\shift}
\shift=0.4cm
\newcommand{\fg}[3]
{
\begin{figure}[ht]
\vspace*{-0cm}

\[
\includegraphics[width=\figwidth]{#1}
\]
\vspace*{\shift}
\caption{\label{#2}
\small
#3
}
\end{figure}}

\newcommand \bea {\begin{eqnarray} }
\newcommand \eea {\end{eqnarray}}

\title{Many-Body Physics: Unfinished Revolution}

\author{Piers Coleman}
\maketitle

\begin{abstract}The study of many-body physics 
has provided a scientific playground of surprise and
continuing revolution over the past half century.  The serendipitous
discovery of new states and properties of matter, phenomena such as
superfluidity, the Meissner, the Kondo and the fractional quantum hall
effects, have driven the development of new conceptual frameworks for
our understanding about collective behavior, the ramifications of
which have spread far beyond the confines of terrestrial condensed
matter physics- to cosmology, nuclear and  particle physics. 
Here I shall selectively review some of
the developments in this field, from the cold-war period, until the
present day. I describe how, with  the
discovery of new classes of collective order, the unfolding
puzzles of high temperature superconductivity and quantum criticality,
the prospects for major conceptual discoveries remain as bright today as they were
more than half a century ago.
\end{abstract}


\section{Emergent Matter: a new Frontier}

Since the time of the Greeks, scholars have pondered over the
principles that govern the universe on its tiniest and most vast
scales. The icons that exemplify these frontiers are very well known -
the swirling galaxy denoting the cosmos and the massive accelerators
used to probe matter at successively smaller scales- from the atom 
down to the quark and beyond. These traditional frontiers
of physics are largely concerned with reductionism: the notion that once
we know the laws of nature that operate on the smallest possible
scales, the mysteries of the
universe will finally be revealed to us\cite{weinberg}.

Over the last century and a half, a period that stretches back
to Darwin and Boltzmann- scientists
have also become fascinated by another notion: the idea that
to understand nature, one also needs to understand and
study the principles that govern collective behavior of vast assemblies of matter. For a
wide range of purposes,  we already know the microscopic laws that
govern matter on the tiniest scales.  For example, a gold atom can be
completely understood with the Schr{\"o}dinger equation and the laws of
quantum mechanics established more than seventy years ago.  Yet, a
gold atom is spherical and featureless- quite unlike the lustrous
malleable and conducting metal which human society so prizes. To
understand how crystalline assemblies of gold atoms acquire the
properties of metallic gold, we need new principles-- principles that
describe the collective behavior of matter when humungous
numbers of gold atoms congregate to form a metallic crystal. 
It is the search for these new principles that defines the frontiers 
of many-body physics in the realms of condensed matter physics and its closely
related discipline of statistical mechanics.

In this informal article, I shall talk about the evolution
of our ideas about the collective behavior of matter since the advent
of quantum mechanics, hoping 
to give a sense of how often unexpected
experimental discovery has seeded the growth of conceptually new ideas
about collective matter. Given the brevity of the article, I 
must apologize for the necessarily selective nature of this discussion.
In particular, I have had to make a painful
decision to leave out a discussion of the many-body physics of
localization and that of spin glasses. I do hope future articles will have opportunity to redress this imbalance.

The past seventy years of development in many-body physics
has seen a  period of 
unprecedented conceptual and
intellectual development. Experimental discoveries of remarkable new
phenomena, such as superconductivity, superfluidity, criticality, liquid
crystals, anomalous metals, antiferromagnetism
and the quantized Hall effect, have
each prompted a renaissance in areas once thought to be 
closed to further fruitful intellectual study. 
Indeed, the history of the field is marked by the most
wonderful and unexpected shifts in perspective and understanding that 
have involved  close linkages between experiment, new mathematics and
new concepts. 

I shall discuss three eras:- the 
immediate aftermath of quantum mechanics---\  many-body physics in the
cold war,  and the modern era of correlated matter
physics. Over this period, physicists' view of the matter 
has evolved \break dramatically- as witnessed by the evolution  in our view of ``electricity''
from the idea of the degenerate electron gas, to the
concept of the Fermi liquid, to new kinds of electron fluid, such as a
the Luttinger liquid or fractional quantum Hall state. Progress was
not smooth and gradual, but often involved the agony, despair and
controversy of the creative process. 
Even the notion that an electron is a 
fermion was controversial.  Wolfgang Pauli, inventor of the 
exclusion principle 
could not initially envisage that this principle 
would apply beyond the atom to macroscopically vast assemblies of
degenerate electrons; indeed, he initially preferred the idea that
electrons were bosons.  Pauli arrived 
at the realization that the electron fluid is a
degenerate Fermi gas with great reluctance, and at
the end of 1925\cite{pauli} gave way, 
writing in a short note to Schr{\"o}dinger that read
\begin{quotation}
\sl ``With a heavy heart, I have decided that Fermi Dirac, not Einstein is
the correct statistics, and I have decided to write a short note on paramagnetism.''
\end{quotation}
\noindent Wolfgang Pauli, letter to Schr{\"o}dinger, November 1925\cite{pauli}.


\shift=-0.2cm
\begin{center}
\figwidth=0.8\textwidth \fg{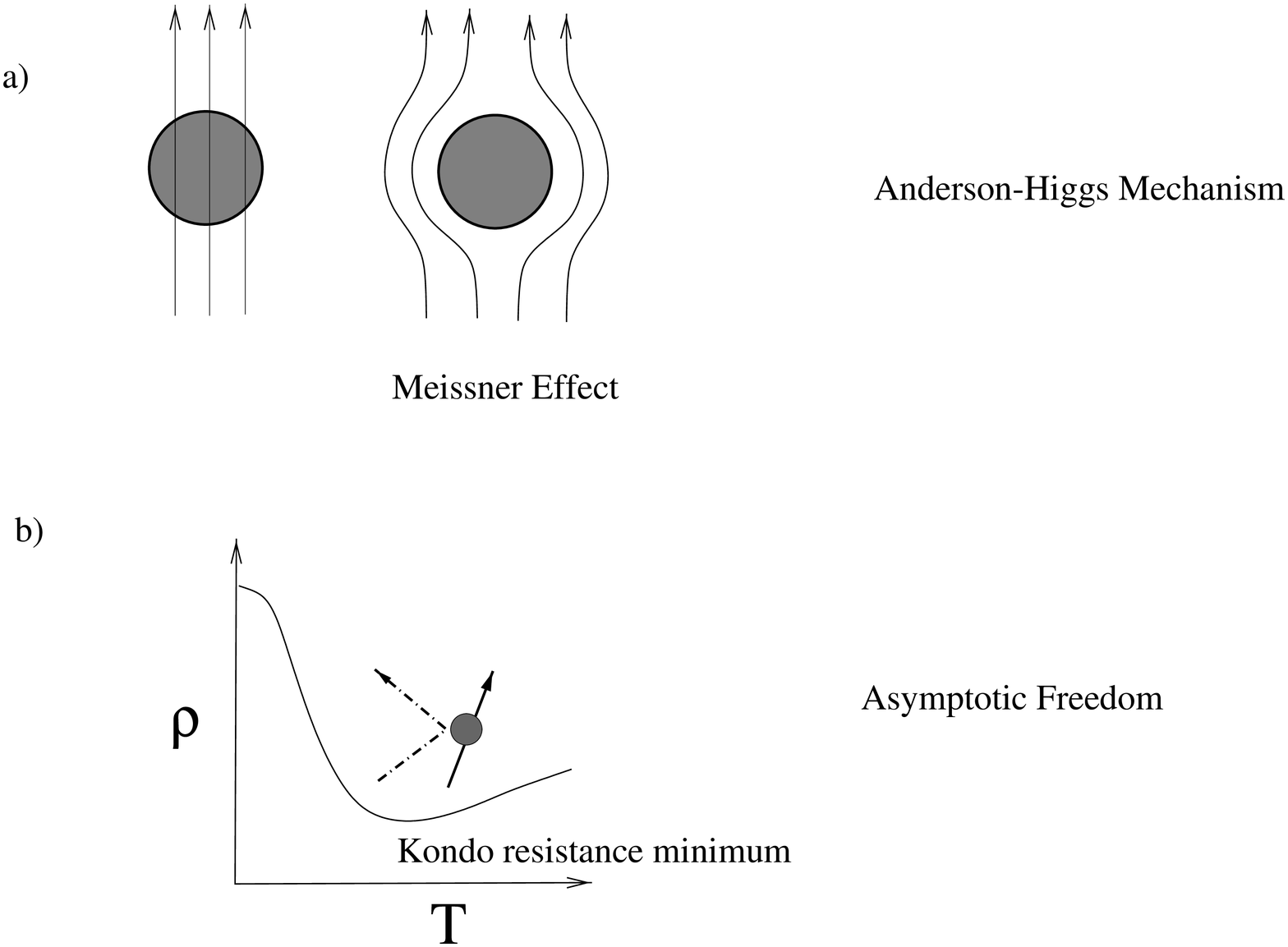}{fig1}{Two 
mysteries of the early
era, whose ultimate resolution 30 years later linked 
them to profound
new concepts about nature. (a) The Meissner effect, 
whose ultimate resolution
led to an understanding of superconductivity and the discovery of the Anderson-Higgs mechanism,  (b) The Kondo resistance minimum, which is linked to
the physics of confinement. 
} 
\end{center}
\shift=-0.9cm

\section{Unsolved riddles of the \hbox{1930s}}

The  period of condensed matter physics between the two world-wars
was characterized by a
long list of unsolved mysteries in the area of magnetism and\break
superconductivity\cite{hoddeson}. Ferromagnetism had
emerged as a shining triumph of the application of quantum
mechanics to condensed matter.  So rapid was the progress in this
direction, that 
N{\'e}el and Landau quickly went on to generalize the idea, predicting the possibility
of staggered magnetism, or antiferromagnetism in 1933\cite{afm}. In a situation
with many parallels today, the experimental tools required to
realize the predicted phenomenon, had to await two decades, for
the development of neutron diffraction\cite{neutrons}.  
During this period, Landau became pessimistic and
came to the conclusion that quantum fluctuations would most
probably destroy antiferromagnetism, as they do in the
antiferromagnetic 1D Bethe chain
 - encouraging one of his students,
Pomeranchuk, to explore the 
idea that spin systems  behave as neutral fluids of fermions\cite{pomeranchuk}.

By contrast, superconductivity remained unyielding
to the efforts of the finest minds in quantum mechanics during the
heady early days of quantum mechanics in the 1920s, a failure derived
in part from a deadly early misconception about
superconductivity\cite{hoddeson}.  It
was not until 1933 that a missing element in the puzzle came to
light, with the Meissner and Ochensfeld discovery that superconductors
are not perfect conductors, but perfect diamagnets.\cite{meissner}
It is this key discovery that led the London brothers\cite{london}
to link superconductivity to a concept of ``rigidity'' in
the many-body electron wavefunction, a notion that Landau and Ginzburg
were to later incorporate in their order parameter treatment of
superconductivity\cite{ginzburg}.  

Another experimental mystery of the 1930's, was the observation of a
mysterious ``resistance minimum'' in the temperature dependent
resistance of copper, gold, silver and other
metals\cite{resistanceminimum}. It took 25 more years for the
community to link this pervasive phenomenon with tiny concentrations
of atomic size magnetic impurities- and another 15 more years to solve
the phenomenon - now known as the Kondo effect- using the concepts of 
renormalization.

\section{Many-Body Physics in the Cold War}

\subsection{Physics without Feynman diagrams}

Many-body physics blossomed after the end of the second world war,  
and as the political walls between the east and west grew with the
beginning of the cold war, a most wonderful period of scientific and
conceptual development, with a frequent exchange of new ideas across
the iron curtain, came into being.  Surprisingly, the Feynman diagram
did not really enter many-body physics until the early 60s, yet
without Feynman diagrams, the many-body community made a sequence of
astonishing advances in the 1950's\cite{pinesmb}.

\begin{figure}[ht]
\centerline{\epsfig{figure=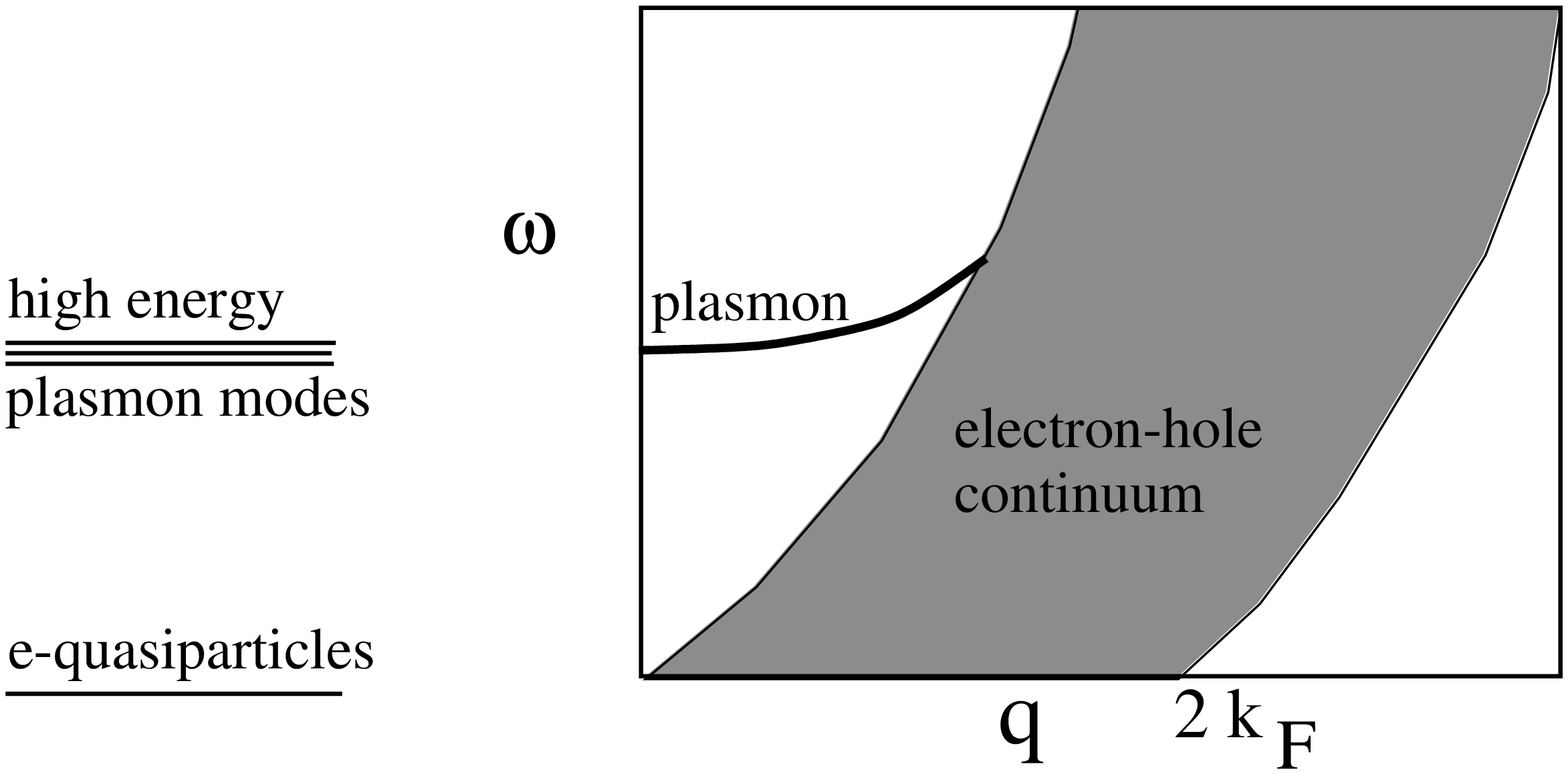,width=8cm}}
\caption{\small Illustrating the Pines-Bohm
idea, that the physics of the electron fluid can be divided up into
high energy collective ``plasmon modes'' and low energy electron quasiparticles.}
\label{fig2}
\end{figure}


The early 1950s saw the first appreciation by the community of the
importance of collective modes.  One of the great mysteries 
was why the non-interacting Sommerfeld model of the
electron fluid worked so well, despite the presence of interactions that are comparable
to the kinetic energy of the electrons.  
In a landmark early paper,  David Bohm and his graduate student, David
Pines\cite{bohm53} realized that they could separate the strongly
interacting gas via a unitary transformation 
into two well-separated sets of excitations- 
a high energy collective oscillations of the  electron
gas, called  plasmons, and low energy electrons. The Pines-Bohm paper
is a progenitor of the idea of renormalization: the
idea that high energy modes of the system can be successively
eliminated to give rise to a renormalized picture of the residual low energy
excitations. 

Feynman diagrams entered many-body physics in the late
1950s\cite{pinesmb}.  The first  applications of the formalism of quantum field 
theory to the many-body physics of bulk electronic matter, made by
Brueckner\cite{earlybrueckner}, were closely followed by Goldstone
and Hubbard's elegant re-derivations of the method using Feynman diagrams
\cite{goldstone57,hubbard57}.  A flurry of activity followed: 
Gell-Mann and Brueckner used the newly discovered ``linked cluster
theorem'' to calculate the correlation energy of the high density
electron gas\cite{brueckner57}, and Galitskii and
Migdal\cite{galitskiimigdal58,galitskii58} in the USSR applied the methods to
the spectrum of the interacting electron gas. 
Around the same
time, Edwards\cite{edwards58} made the first applications of Feynman's methods to
the problem of elastic scattering off disorder. 

One of the great theoretical leaps of this early period was
the invention of the concept of imaginary time\footnote{The key ideas
of the imaginary time approach were certainly known to Kubo prior to the
first publication by Matsubara. 
P. W. Anderson recalls being shown the key ideas of this technique,
including the antiperiodicity of the Fermi Green function,  by Kubo,
Matsubara's mentor, in 1954. }.  
 The earliest
published discussion of this idea occurs in the papers of
Matsubara\cite{matsubara}.  Matsubara 
noted the remarkable similarity between the time evolution operator of quantum mechanics
\bea
U(t)= e^{-i t H/\hbar}
\eea
and the Boltzmann density matrix
\bea
\rho(\beta) = e ^{- \beta H} = U ( - i \hbar \beta), 
\eea
where $\beta = 1/(k_B T)$ and $k_B$ is Boltzmann's constant. 
This parallel suggested that one could convert conventional quantum
mechanics into finite temperature
quantum statistical mechanics by using a time-evolution operator where
real time is replaced by 
imaginary time, 
\bea
t \rightarrow - i \tau \hbar.
\eea
Matsubara's ideas took a further leap into the realm of the practical, when\break
Abrikosov, Gorkov and Dzyaloshinski (AGD) 
\cite{agd} 
showed that the method was dramatically simplified by Fourier
transforming the imaginary time electron Green function into the
frequency domain. They noted for the first time that the antiperiodicity of the Green function $G (\tau
+\beta )= -G (\tau )$ meant that  the continuous frequencies of zero temperature
physics are replaced by the discrete frequencies
$\omega_{n}= (2n +1 ) \pi T$, that we now call
the ``Matsubara
frequency''. In their paper, the finite temperature propagator
\bea
G (\omega_{n},\vec{p}) = [ i \omega_{n}+\mu - \epsilon (\vec{ p})]^{-1}
\eea
for the electron makes its first appearance. 

Another great conceptual leap of the early cold war, was the
development of the concept of the ``elementary excitation'', or
``quasiparticle'', as a way to understand the low-energy excitations
of many-body systems.  The idea of a quasiparticle is usually
associated with Landau's pioneering work on the Fermi liquid, which appeared in
1957. The basic concept of elementary excitation appears to
have been in circulation on both sides of the Iron Curtain throughout
much of the fifties. 
The term ``quasiparticle'' certainly  appears in Boguilubov's\cite{boguilubov}
paper on the theory of superfluidity in 1947.
However, Landau's work on Fermi liquids certainly added tremendous
clarity to the quasiparticle idea. 
Landau\cite{landaufl}, stimulated by early
measurements on liquid He-3, realized that interacting fermi gases
could be understood with the concept of ``adiabaticity''- the notion that
when interactions  are turned on adiabatically,  the original single-particle
excitations of the  Fermi liquid, evolve without changing their charge
or spin quantum numbers, into ``quasiparticle'' excitations of the interacting
system.  Today, Landau's Fermi liquid theory is the foundation for
the modern ``standard model'' of the electron fluid. 

\subsection{Broken Symmetry}

Two monumental achievements of the cold-war era  deserve
separate mention: the discovery of ``broken symmetry'' and the renormalization group. 
In 1937, Landau\cite{landau37} formulated the concept of broken symmetry-
proposing that phase transitions take place via the process of symmetry
reduction, which he described in terms of his order parameter concept. 
In the early fifties, Onsager and Penrose\cite{odlro},  refined Landau's concept of broken symmetry to propose
that  superfluidity could be understood
as a state of matter in which the two-particle density matrix 
\bea
\rho(r,r')= \langle \hat \psi \dg(r)\hat \psi(r')\rangle
\eea
can be factorized:
\bea
\rho(r,r') = \psi^*(r)\psi(r) + \hbox{small terms}
\eea
where
\bea
\psi(r)= \sqrt{\rho_s} e^{i \phi} = \langle N-1 \vert \hat \psi(r) \vert N \rangle.
\eea
is the order parameter of the superfluid, 
$\rho_s$ is the superfluid density and $\phi$ the phase of the condensate.
This concept of ``off-diagonal long-range order''  later became generalized to fermi systems
as part of the BCS theory of superconductivity\cite{bcs,gorkov58}, where the off-diagonal
order parameter
\bea
F(x-x') = \langle N-2\vert \hat \psi_{\downarrow}(x) \hat \psi_{\uparrow} (x')\vert N\rangle,
\eea
defines the wavefunction of the Cooper pair.  

Part of the inspiration for a state with off-diagonal long-range
order in BCS theory came from work by 
Tomonaga\cite{tomonaga} involving a pion condensate around the
nucleus.  
Bob Schrieffer wrote down the BCS wavefunction while
attending a many-body physics meeting in 1956 at the Stephens Institute
of Technology, in New Jersey. In a recollection he
writes\cite{bobschrieffer}
\begin{quotation}
{\sl ``While attending that meeting it occurred to me that because of the
strong overlap of pairs perhaps a statistical approach analogous to a
type of mean field would be appropriate to the problem. Thinking back
to a paper by Sin-itiro Tomonaga that described the pion cloud around
a static nucleon \cite{tomonaga}, I tried a ground-state wave function 
$\vert \psi _{0}\rangle $ written as
\begin{equation}\label{}
\vert \psi _{0}\rangle  = \prod _{k}\left(u_{k}+v_{k}c\dg
_{k\uparrow}c\dg _{-k\downarrow } \right)\vert 0\rangle 
\end{equation}
where $c\dg _{k\uparrow}$ is the creation operator for an electron
with momentum $k$ and spin up, $\vert 0\rangle $ is the vacuum
state, and the amplitudes $u_{k}$ and $v_{k}$ are to be determined''.}
\end{quotation}

One of the remarkable spin-offs of superconductivity, was that it led
to an understanding of how a gauge boson can acquire a mass as a
result of symmetry breaking.  This idea was first discussed by
Anderson in 1959\cite{anderson}, and in more detail in
1964\cite{andersonhiggs,andersonhet}, but the concept evolved further
and spread from Bell Laboratories to the particle physics community,
ultimately re-appearing as the Higg's mechanism for spontaneous
symmetry breaking in a Yang Mills theory. The Anderson-Higgs mechanism
is a beautiful example of how the study of cryogenics led to a
fundamentally new way of viewing the universe, providing a mechanism
for the symmetry breaking between the electrical and weak forces in
nature.  

Another consequence of broken symmetry concept is the notion
of ``generalized rigidity''\cite{basicnotions}, a concept which has its
origins in London's early model of superconductivity\cite{london}
and the two-fluid models of superfluidity proposed independently by
Tisza\cite{tisza} and Landau\cite{landausf}, according to which,
if the phase of a boson or Cooper pair develops a rigidity, then it
costs a phase bending energy \bea U(x)\sim \frac{1}{2}\rho_s (\nabla
\phi(x))^2 , \eea from which we derive that the ``superflow'' of
particles is directly proportional to the amount of phase bending, or the
gradient of the phase \bea j_{\hbox{s}}= \rho_s \nabla \phi.  \eea
Anderson noted\cite{basicnotions} that we can generalize this concept
to a wide variety of broken symmetries, each with their own type of
superflow (see table 1).  Thus broken translation symmetry leads to
the superflow of momentum, or sheer stress, broken spin symmetry leads
to the superflow of spin or spin superflow. There are undoubtedly new
classes of broken symmetry yet to be discovered.

\vfill \eject 

\input{table1x.tex}

\subsection{Renormalization group}

The theory of second order phase
transitions was studied by Van der Waals in the 19th century, and thought
to be  a closed field\cite{domb}.  Two events- the experimental observation of critical
exponents  that did not fit the predictions of mean-field theory\cite{guggenheim,voronel}, and
the solution to the 2D Ising model\cite{onsager}, forced condensed matter physicists
to revisit an area once thought to be closed.  The revolution that
ensued literally shook physics from end to end, furnishing us with
a spectrum of new concepts and terms, such as\begin{itemize}

\item scaling theory\cite{widom,kadanoff,pokrovsky}, 
\item universality- the idea that the essential physics at long length scales
is independent of all but a handful of short-distance details, such
as the dimensionality of space and the symmetry of the order parameter. 
\item renormalization- the process by which short-distance, high energy
physics is absorbed by adjusting the parameters inside the Lagrangian or Hamiltonian.
\item fixed points- the limiting form of the Lagrangian or Hamiltonian
as short-distance, high energy physics is removed
\item running coupling constant-- a coupling constant whose magnitude
changes with distance,
\item upper critical dimensionality- the dimension above which mean-field
theory is valid. 

\end{itemize}that appeared as part of the new 
``renormalization group''\cite{migdal,kadanoff2,fisher,wilson}. 
The understanding of classical phase transitions required the remarkable fusion of
universality, together with the new concepts of
scaling, renormalization and the application of tools borrowed from quantum field theory.
These developments are a main stay of modern
theoretical physics, and their influence is felt far outside the realms
of condensed matter. 

One of the unexpected dividends of the renormalization
group concept, in the realm of many-body theory, was the solution of the Kondo
effect: the condensed matter analog of quark confinement.  
By the late fifties, the resistance minima in copper, gold and silver alloys
that had been observed since the 1930s\cite{resistanceminimum}, 
had been identified with magnetic impurities, but the mechanism for
the minimum was still unknown. 
In the early
60's, Jun Kondo\cite{kondo} was able to identify this resistance
minimum, as a consequence of antiferromagnetic interactions
between the local moments and the surrounding electron gas. 
The key ingredient
in the Kondo model, is an antiferromagnetic interaction between a local moment
and the conduction sea, denoted by
\bea
H_I = J \vec \sigma(0)\cdot \vec S
\eea
$\vec S$ is a spin $1/2$ and  $\vec \sigma(0)$ is the spin density
of the conduction electrons at the origin. 
Kondo\cite{kondo} found that when he calculated the scattering rate $\tau^{-1}$
of electrons off a magnetic moment to one order higher than Born
approximation, 
\bea
\frac{1}{\tau } \propto [
J\rho + 2 (J\rho
)^{2 } \ln \frac{D}{T}
]^{2},
\eea
where $\rho$ is the
density of state of electrons in the conduction sea and $D$ is the
width of the electron band. As the temperature is lowered, the
logarithmic term grows, and the scattering rate and resistivity
ultimately rises, connecting the resistance minimum with the antiferromagnetic
interaction between spins and their surroundings. 

A deeper understanding of this logarithm
required the renormalization group
concept\cite{andersonyuval,wilson,nozieres}. 
By systematically taking the effects of  high
frequency virtual spin fluctuations into account, it became clear that
the bare coupling $J$ is replaced by a renormalized quantity 
\bea
J\rho 
(\Lambda)= J\rho + 2 (J\rho
)^{2 } \ln \frac{D}{\Lambda}
\eea
that depends on the scale $\Lambda$ of the cutoff, so that the
scattering rate is merely given by $1/\tau \propto (\rho J
(\Lambda))^{2}\vert _{\Lambda\sim T}$. The corresponding 
renormalization equation 
\bea
\frac{\partial J\rho }{\partial \ln
\Lambda} = \beta(J\rho )= - 2 ( J\rho) ^2 + O(J^3) 
\eea
contains a ``negative $\beta $ function'': the hallmark of a coupling
which  dies away at high energies (asymptotic freedom), but which grows
at low energies, 
ultimately reaching  a value of order unity when the
characteristic cut-off is reduced to the scale of the so called
``Kondo temperature'' $T_K \sim D e^{- 1/J}$. 

The ``Kondo'' effect is a manifestation of the phenomenon of ``asymptotic freedom'' that also governs quark physics.
Like the quark, at high energies the local moments inside metals 
are asymptotically free, but
at energies below the Kondo temperature, they interact so strongly
with the surrounding electrons 
that they become screened or
``confined'' at low energies, ultimately forming a Landau Fermi
liquid\cite{nozieres}.
It is a remarkable that the
latent physics of confinement, hiding within cryostats 
in the guise of the Kondo
resistance minimum, remained a mystery for  more than 40 years,
pending purer materials, the concept of local moments and the
discovery of  the renormalization group.

\subsection{The concept of Emergence}

The end of the cold-war period in many-body physics is 
marked
by Anderson's statement of the concept of emergence. 
In a short 
paper, originally presented as part of a Regent's lecture entitled
``More is different'' at San Diego
in the early seventies\cite{emergence}, Anderson defined the concept of emergence with the 
now famous quote

\begin{quotation}
\sl ``at each new level of complexity, entirely new properties appear,
and the understanding of these behaviors requires research which I think is
as fundamental in its nature as any other.''
\end{quotation}

Anderson's quote underpins a modern attitude to condensed matter physics-
the notion that the study of the collective principles that govern
matter is a frontier unto itself, complimentary, yet separate to those
of cosmology, particle physics and biology. 

\section{Condensed Matter Physics in the New Era}

\subsection{New States of Matter}

By the end of the 1970's few condensed matter physicists had 
really internalized the consequences of emergence. In the early eighties,
most members of the community were for the most part, 
content with a comfortable notion that the principle constraints on
the behavior and possible ground-states of dense
matter were already known. Superconductivity was widely believed to be limitedto  below about
25K\cite{mcmillan}.  The ``vacuum'' state of metallic behavior was firmly
believed to be the Landau Fermi liquid, and no significant departures
were envisaged outside the realm of one-dimensional conductors.
Tiny amounts of magnetic impurities were known to be anathema to
superconductivity. These principles were so entrenched in the
community that the first observation\cite{bucher} of heavy electron
superconductivity in the magnetic metal UBe$_{\rm 13}$ was mis-identified
as an artifact, delaying acceptance of this phenomenon by another decade.
By the end of the 80's all of these popularly held
principles had been exploded by an unexpected sequence of discoveries,
in the areas of heavy electron physics, the quantum Hall effect and
the discovery of high temperature superconductivity. 
\begin{figure}[ht]
\centerline{\epsfig{figure=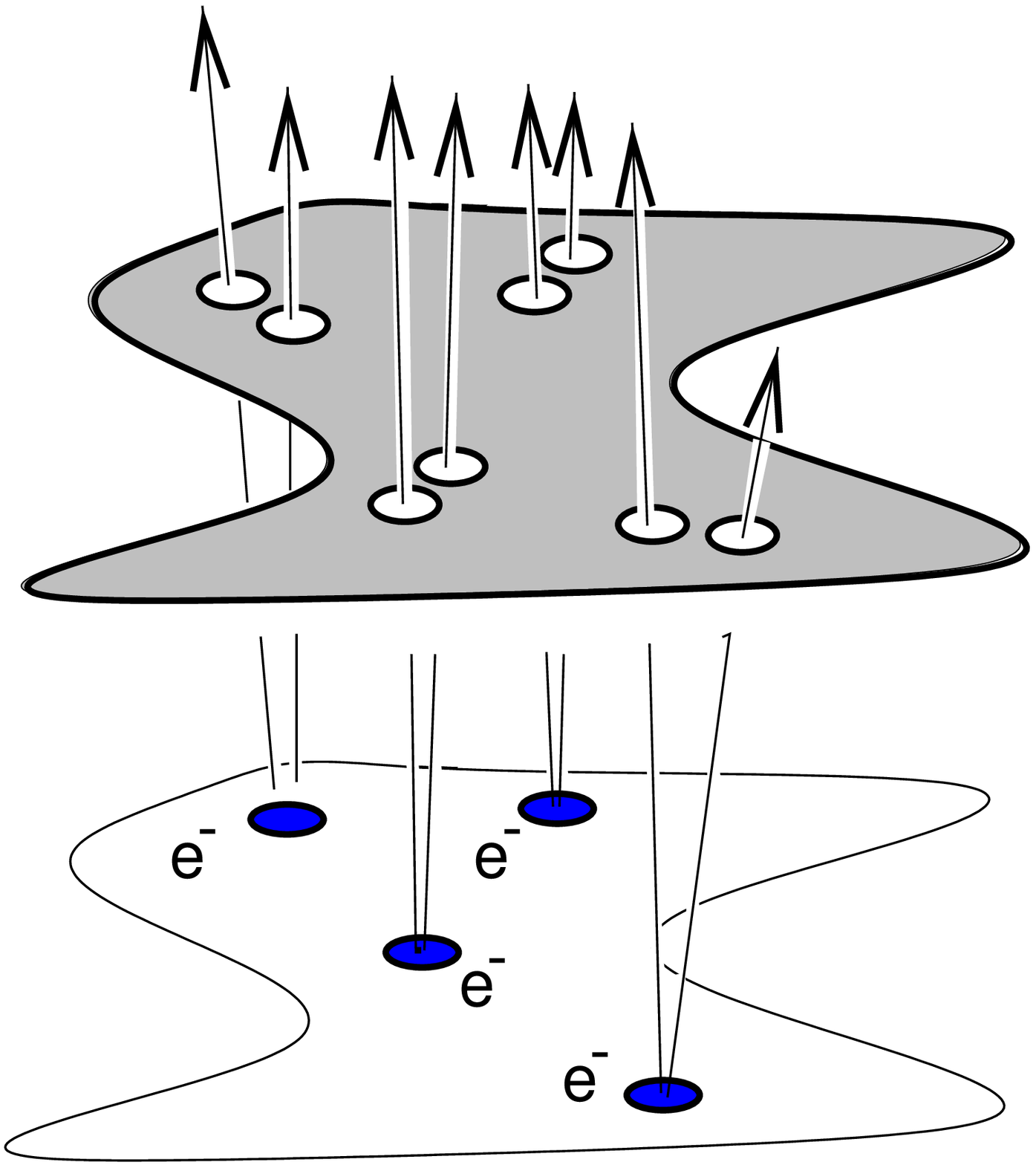,width=5cm}}
\caption{\small Illustrating the binding of two vortices to each electron, to form the
$\nu =1/3$ Laughlin ground-state.}
\label{fig3}
\end{figure}


\subsubsection{Fractional Quantum Hall Effect}

In the 1930's Landau had discussed the quantum mechanics of electron motion
in a magnetic field\cite{landau30}, predicting the quantization of
electron kinetic energy 
into discrete Landau levels 
\bea
\frac{\hbar^2  (k_x^2 + k_y^2)}{2m}\rightarrow
\frac{\hbar eB}{m}(n+\frac{1}{2}), \qquad (n=0,\ 1,\ 2 \dots ).
\eea
Landau quantization had been confirmed in metals, where it produces
oscillations in the field-dependent resistivity (Shubnikov de Haas oscillations)
and magnetization (de Haas van Alphen oscillations), and the field was thought
mature. In  the seventies, advances in semiconductor
technology and the availability of high magnetic fields, 
made it possible to examine two dimensional electron fluids 
at high fields, when the spacing of the Landau levels is 
so large that the electrons drop into the lowest Landau level, so that 
their dynamics is entirely dominated by mutual Coulomb interactions. 
Remarkably,  the Hall constant of these electron fluids 
was found to be quantized with values
$
R_H = \frac{V} {I} = \frac{h}{\nu e^2}, 
$
where at lower fields, $\nu = 1, 2, 3\dots $ is an integer, but at higher fields,
$\nu$ acquires a fractionally quantized values $\nu = 1/3,\ 1/5, \ 1/7
\dots $.  
Laughlin\cite{bob2} showed that the fractional quantum Hall effect 
is produced by interactions,
which stabilize a new 
type of
electron fluid where the Landau level has fractional  filling factor
$\nu = 1/ (2M+1)$.
In Laughlin's approach, 
the  electron fluid is pierced by 
``vortices'' which identify zeroes in the electron wavefunction. 
Laughlin proposed that electrons bind to these vortices 
to avoid other electrons, and he incorporated this
physics into his celebrated wavefunction by attaching each electron to an 
even number $2M$ of
vortices.
\bea
\Psi(\{z_i\}) = \prod_{i>j} (z_i - z_j)^{2M+1} \exp\left[
\sum_i \vert z_i\vert^{2}/4 l_o^2
\right]
\eea
where $l_o = \sqrt{\hbar/e B}$ is the magnetic length. 
The excitations in this state are gapped, with both fractional
charge and fractional statistics: an entirely new electronic ground
state. Moreover, the wavefunction is robust against the 
details of the Hamiltonian from which it is derived. 

This break-through opened  an entire field of investigation\cite{fqhe,murthy} into the new world
of highly correlated electron physics,  bringing  a whole range of new
concepts and language, such as
\begin{itemize}

\item fractional statistics quasiparticles- 

\item composite fermions- 

\item Chern-Simons terms.

\end{itemize}
Equally importantly, the fractional quantum Hall effect made the
community poignantly aware of the profound transformations that become
possible in electronic matter when the strength of interactions
becomes comparable to, or greater than the kinetic energy.

\subsubsection{Heavy Electron Physics}

The discovery of heavy electron materials in the late
seventies\cite{ott,steglich}
forced condensed matter
physicists to severely
revise their understanding about how local moments interact with the electron fluid.
In the late seventies, electron behavior in metals was neatly categorized into 
\begin{enumerate}

\item ``delocalized'' behavior, where
electrons form Bloch waves, and 

\item ``localized''  behavior,
where the electrons in question are bound near a particular
atom in the material. Such unpaired spins 
form tiny atomic magnets called ``local moments'' that tend to align
at low temperatures and are extremely damaging to superconductivity.

\end{enumerate}
Heavy fermion metals
completely defy these norms, for they contain a
dense array of magnetic  moments, yet
instead of magnetically ordering the moments develop a highly correlated
paramagnetic ground-state with the
conduction electrons. When this happens, the resistivity of the
metal drops abruptly,  forming  a highly correlated Landau Fermi liquid
in which electron masses rise in excess of 100
times the bare electron mass\cite{hewson}. 

Heavy electron physics is, in essence the direct descendant of
the resistance minimum physics first observed in simple metals in the early
1930's.  Our current understanding of heavy fermions is based on the
notion, due to Doniach\cite{doniach}, that the ``Kondo effect'' seen for individual
magnetic moments, survives inside the dense magnetic arrays of 
heavy fermion compounds to produce the heavy fermion
state. The heavy electrons that propagate in these materials are
really the direct analogs of nucleons formed from confined quarks.
Curiously, one of the most useful theoretical methods for describing these systems
was borrowed from particle physics.  Heavy electrons are formed in
$f$-orbitals 
which are spin-orbit coupled with a  large spin  degeneracy $N=2j+1$.  One of the most useful methods for developing a mean-field
description of the heavy electron metal is the $1/N$ expansion, inspired
by analogies with the $1/N$ expansion in the spherical model of statistical
mechanics\cite{berlin} and the 
$1/N$ expansion in the number of colors in
Quantum Chromodynamics\cite{witten,read,coleman,auerbach}.
Here the basic idea is that $1/N$ plays the role of an effective
Planck's constant 
\bea\label{}
\frac{1}{N} \sim \hbar_{eff},
\eea
so that as $N\rightarrow \infty$, certain operators, or combinations
of operators in the Hamiltonian behave as new classical variables. The
physics can then be solved in the large $N$ limit as a special kind of
classical physics, and the corrections to this limit are then expanded
in powers of $1/N$.  In this way much of the essential physics of the
heavy electron paramagnet is captured as a semi-classical expansion
around a new class of mean-field theory, where the width of the heavy
electron band plays the role of an order parameter.

\subsubsection{High Temperature Superconductivity}

The discovery of high temperature superconductivity, with transition temperatures
that have spiraled way above the theoretically predicted maximum possible transition temperatures,
to its current maximum of 165K, stunned the physics community. 
These systems are formed by  adding charge
to an insulating state where electrons are localized in an antiferromagnetic
array. 
Several aspects of these materials radically challenge our understanding of correlated electron systems,
in particular:
\begin{itemize}

\item The close vicinity between insulating and superconducting
behavior in the phase diagram, which suggests
that the insulator and superconductor may derive from closely related
ground-state wavefunctions\cite{science,so5}.

\item The ``strange metal'' behavior of the optimally doped materials.
Many properties of this state tell us that it is not a Landau Fermi liquid, such as
the linear resistivity 
\bea
\rho = \rho_o + A T
\eea
extending from the transition temperature, up to the melting temperature.  This linear resistivity
is known to originate in an electron-electron scattering rate
$\Gamma(T) \sim  k_{B}T$, 
that grows linearly with temperature, 
which has been called a ``marginal Fermi liquid'' \cite{marginal}. 
In conventional metals, the inelastic scattering rate  grows quadratically
with temperature. Despite 15 years of effort, the origin of the 
linearity of the scattering rate remains a mystery.

\item The origin of the growth of a pseudogap in the electron spectrum for ``under-doped'' superconductors. 
This soft gap in the excitation spectrum signals the growth of correlations amongst the electrons prior
to superconductivity, and some believe that it signals the formation of pairs, without coherence\cite{kivelson95}.  

\end{itemize}
The radical simplicity of many of the properties of the cuprate
superconductors leads many to believe that their ultimate
solution will require a conceptually new description
of the interacting electron fluid. 
\begin{figure}[ht]
\centerline{\epsfig{figure=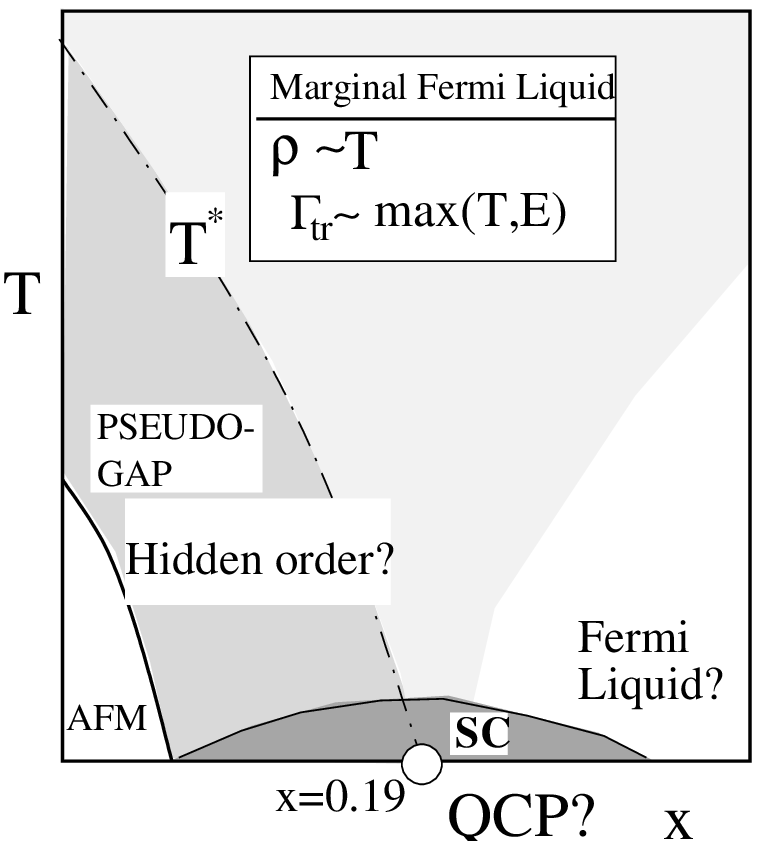,width=6cm}}
\caption{\small Schematic phase diagram for cuprate superconductors, where $x$ is doping and $T$ the temperature,  showing the location of a possible 
quantum critical point.}
\label{fig4}
\end{figure}


The qualitative phase diagram is shown in
 Fig(\ref{fig4}), showing
three distinct regions- the over-doped region, the fan of ``marginal
Fermi liquid behavior'' and the under-doped region. The theoretical study 
of this phase diagram has proven to be a huge engine
for new ideas, such as
\begin{itemize}

\item Spin charge separation- the notion that the spin-charge coupled electron breaks up into
independent collective charge and spin excitations, as in one dimensional fluids. 

\item Hidden order- the notion that the pseudo-gap is a consequence of the formation of an as-yet
unidentified order parameter, such as orbital magnetism ($d$-density waves)\cite{varma,ddw} or stripes\cite{stripes}.

\item Quantum criticality- the notion that the strange-metal phase of the cuprates is a consequence
of a ``quantum critical point'' around a critical doping of about $x_c\sim
0.2$\cite{loram} In this
scenario, the pseudo-gap is associated with the
growth of ``hidden order''  and marginal Fermi liquid behavior is
associated with the quantum fluctuations emanating from the
quantum critical point.

\item Pre-formed pairs- the idea that the under-doped pseudo-gap region of the phase diagram is a consequence of the formation of phase-incoherent pairs which form at the pseudo-gap temperature\cite{kivelson95}. 

\item Resonating Valence Bonds- the idea that superconductivity can be regarded as a fluid of spinless charged holes, moving in a background of singlet spin pairs.\cite{science} 

\item New forms of gauge theory, including $Z_2$\cite{senthil}, $SU(2)$\cite{wen} and even supersymmetric gauge theories\cite{super}
that may describe the manifold of states  that is highly constrained by the strong coulomb interactions
between electrons in the doped Mott insulator. 

\end{itemize}
Many of these ideas enjoy some particular realization in non-cuprate materials, and in this way,
cuprate superconductivity has stimulated a huge growth of new concepts
and ideas in many-body physics.


\subsection{Quantum Criticality}

The concept of quantum criticality: the idea that a zero temperature
phase transition will exhibit critical order parameter fluctuations in
both space and time, was first introduced by John Hertz during the
hey-days of interest in critical phenomena, but was regarded as an
intellectual curiosity.\cite{hertz} Discoveries over the past decade
and a half have revealed the ability of zero-temperature quantum phase
transitions to qualitatively transform the properties of a material at
finite temperatures. For example, high temperature superconductivity
is thought to be born from a new metallic state that develops at a
certain critical doping in copper-perovskite materials.\cite{loram}
Near a quantum phase transition, a material enters a weird state of
``quantum criticality'': a new state of matter where the wavefunction
becomes a fluctuating entangled mixture of the ordered, and disordered
state. The physics that governs this new quantum state of matter
represents a major unsolved challenge to our understanding of
correlated matter.

A quantum critical point (QCP) is a singularity in the phase diagram:
a point $x=x_{c}$ at zero-temperature 
where the
characteristic energy scale $k_{B}T_{o} (x)$ of excitations above the ground-state
goes to zero. (Fig.~5.).\cite{sachdevbook,continentino,moriya,millis,varma2001,pepin} 
The QCP 
affects the broad  wedge of phase
diagram where $T> T_{o} (x)$.  
In this region of the material
phase diagram, the critical quantum fluctuations are cut-off by
thermal fluctuations after a correlation time given by the Heisenberg 
uncertainly principle
\bea
\tau \sim \frac{\hbar }{k_{B}T}.
\eea
As a material is cooled towards a quantum critical point,
the physics probes the critical quantum fluctuations
on longer and longer time-scales. 
Although the ``quantum critical''  region of the phase
diagram where $T>T_{o} (x)$ is  not a strict phase, 
the absence of any scale to the excitations other
than temperature itself qualitatively transforms the properties of the
material in a fashion that we would normally associate with a new
phase of matter.
\begin{figure}[ht]
\centerline{\epsfig{figure=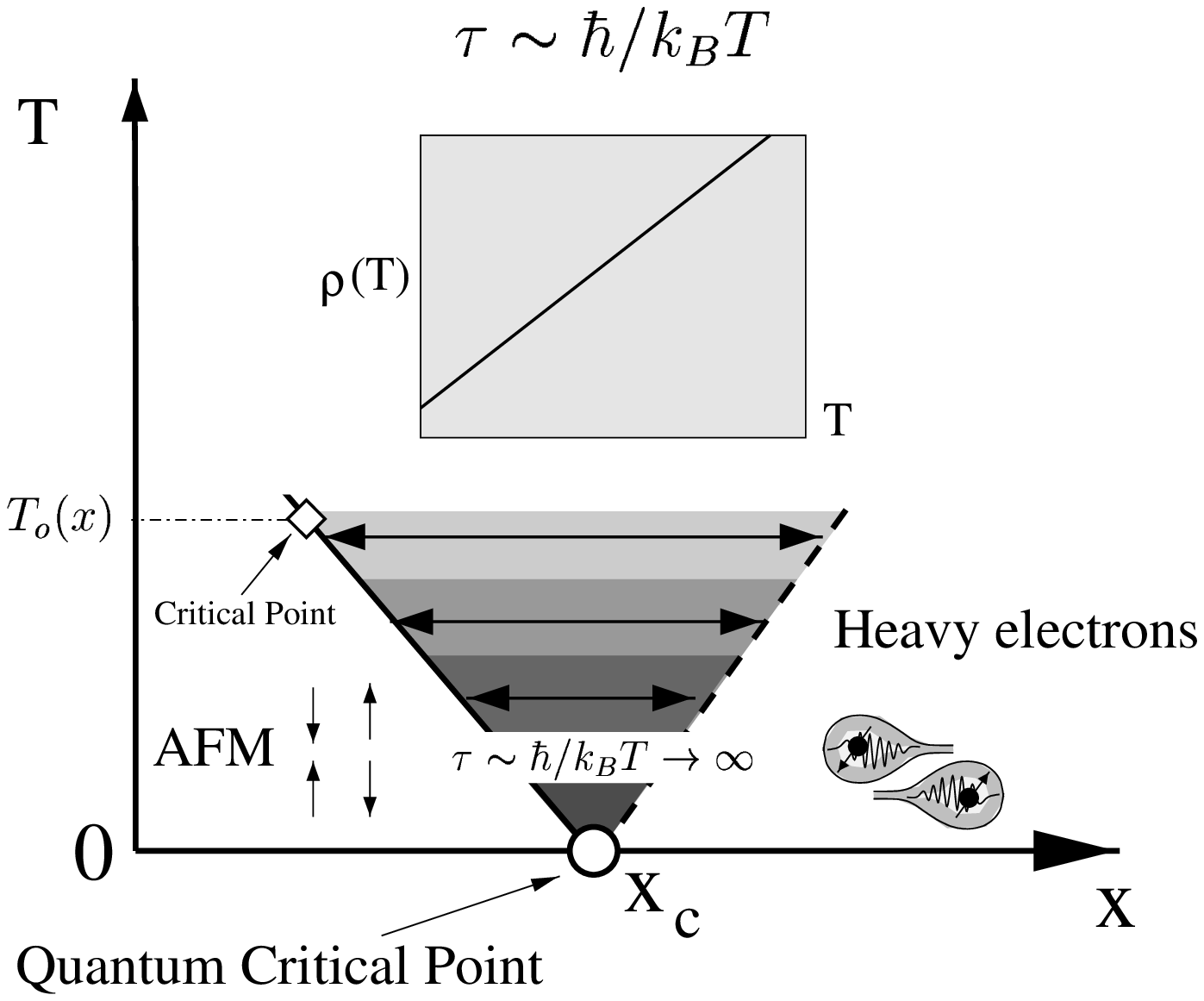,width=8cm}}
\caption{\small Quantum criticality
in heavy electron systems.  For
 $x<x_{c}$ spins become ordered for $T<T_{o} (x)$ forming
an antiferromagnetic Fermi liquid  ; for $x>x_{c}$, 
composite bound-states form between spins and electrons  at $T<T_{0}
(x)$ producing a
heavy Fermi liquid.  ``Non-Fermi liquid behavior'', in which the
characteristic energy
scale is temperature itself, and resistivity is quasi-linear, develops in the wedge shaped region between these two phases. The nature of the critical
Lagrangian governing behavior at $x_c$ is currently a mystery.}
\label{fig5}
\end{figure}
  

Quantum criticality has been extensively studied in heavy electron
materials, in which the antiferromagnetic phase transition temperature
can be tuned to zero by the application of a pressure, field or
chemical doping.  Close to quantum
criticality, these materials exhibit a number of tantalizing similarities
with the cuprate superconductors\cite{pepin}:
\begin{itemize}

\item a predisposition to form
anisotropic superconductors, 

\item the formation of a strange metal with quasi-linear resistivity in the critical region

\item the appearance of temperature as the only scale in the electron excitation spectrum
at criticality, reminiscent of ``marginal Fermi liquid behavior''

\end{itemize}
Hertz proposed that quantum criticality could be understood by
extending classical criticality to order parameter fluctuations in
imaginary time, using a Landau Ginzburg functional that includes the
effects of dissipation:
\begin{equation}
F = \int_0^{1/T} d \tau\int d^dx \left\{
\vert (\nabla+ i {\bf Q}_o )\psi \vert^2 + \xi^{-2} \vert \psi \vert^2 + U \vert \psi\vert ^4
\right\} +F_D
\end{equation}
where ${\bf Q}_o$ is the ordering vector of the antiferromagnet, $\xi$ the correlation length which vanishes at the QCP and 
\bea
F_D = 
\sum_{i\nu_{n}}\int \frac{d^{3}q}{(2\pi)^{3}}\vert \psi
(q,\nu_{n})\vert ^{2}
\frac{\vert
\nu_{n}\vert}{\Gamma_{\bf q}}, \qquad\qquad (\nu = 2 n \pi T)
\eea
is a linear damping rate 
derived from
the density of particle-hole excitations in the Fermi sea. 
An important feature 
of this ``$\phi^4$'' Lagrangian is that the momentum dependence enters
with twice the power of the frequency dependence, 
the time dimension counts as {$z=2$} space dimensions, and 
the effective dimensionality
of the theory is 
\begin{equation}
D= d + z = d+2,
\end{equation}
so that $D=5$ for the three dimensional model, pushing it above
its upper critical dimension. 

In heavy electron
materials, there is a growing sense that the Hertz approach 
can not explain the physics of quantum criticality. Many of the properties
of the QCP, such as the appearance of non-trivial exponents in the
quantum spin correlations, with $T$ as the only energy scale, suggest that
the underlying critical Lagrangian lies {\sl beneath} its critical dimension.
Also, all experiments indicate
that the energy spectrum of the quasiparticles in the Landau Fermi liquid
either side of the QCP, {\sl telescopes to zero}, driving the masses of quasiparticle excitations 
to infinity, and pushing the characteristic Fermi temperature to zero at the quantum critical point.  
Yet the Hertz model predicts that 
most the electron quasiparticle masses should remain finite at an antiferromagnetic QCP.

This has led some 
to propose that unlike classical criticality, we can not use
Landau Ginzburg theory as a starting point for an examination of the
fluctuations: 
{\sl a new mean-field theory} must be found. 
One of the ideas of particular interest, is the
idea of ``local quantum criticality'', whereby the quantum fluctuations of the spins
become critical in time, but not space at a QCP\cite{qmsi}.   
Another idea, is that at a heavy electron quantum critical point,
the heavy electron quasiparticle disintegrates into separate spin and charge degrees of freedom.  Both ideas
require radically new kinds of mean-field theory, raising the prospect
of a discovery of a wholly new class of critical phenomena\cite{pepin}. 

I should add that Chapline and Laughlin have suggested that 
quantum criticality may have cosmological
implications, proposing that the event horizon of a black hole
might be identified with a 
quantum critical interface where the characteristic scales
of particle physics might, in complete analogy with condensed matter, telescope to zero\cite{cosmobob}. 

\begin{figure}[ht]
\centerline{\epsfig{figure=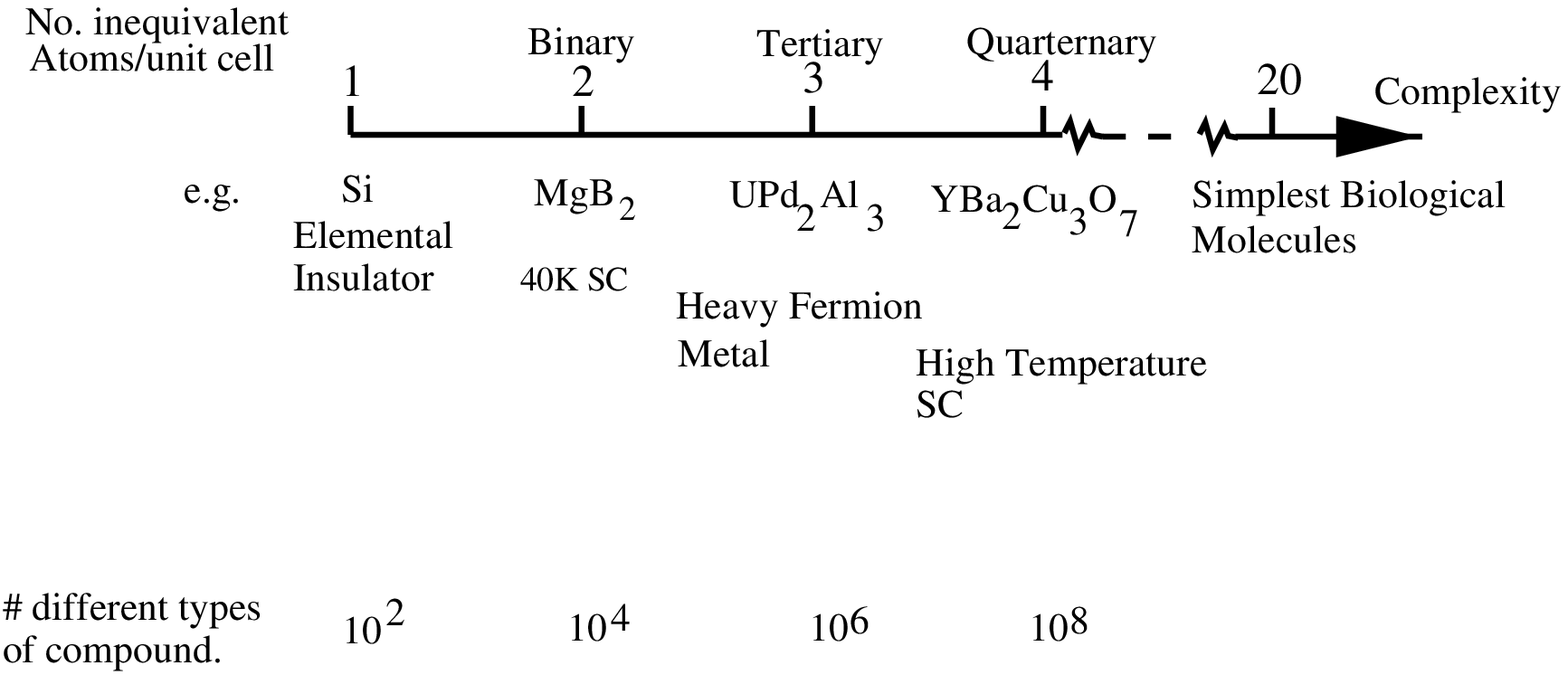,width=10cm}}
\caption{\small The ``axis of complexity''.}
\label{fig6}
\end{figure}


\section{The nature of the Frontier}

This article has tried to illustrate
how condensed matter physics has had a central influence in
the development of our ideas about collective matter, both in the lab,
and on a cosmological scale.  Many simple phenomena seen in the
cryostat, illustrate fundamentally new principles of nature that
recur throughout the cosmos. Thus, the discovery of superconductivity
and the Meissner effect has contributed in a fundamental way to our
understanding of broken symmetry and the Anderson Higg's mechanism. In
a similar way, the observation of the resistance minimum in copper,
provides an elementary example of the physics of confinement, and
required an understanding of the principles of the renormalization
group for its understanding. The interchange between the traditional frontiers-
and the emergent frontier of condensed matter physics is as live
today, as it has been over the past four decades- for example-
insights into conformal field theory gained from the study of 2D phase transitions\cite{cardy} currently play a major role in the description of  D-brane solitons\cite{strominger} in superstring theory. In the future, newly discovered phenomena, such as quantum
criticality are likely to have their cosmological counterparts as well.

One way of visualizing the frontier, is to consider that in
the periodic table, there are about 100 elements.  As we go out along
the complexity axis (Fig. \ref{fig6}), from the elements to the binary, tertiary and
quaternary compounds, the number of possible ordered crystals
exponentiates by at least a factor of 100 at each stage, and with it
grows the
potential for discovery of fundamentally new states of matter. 
Only two years ago- a new high
temperature superconductor MgB$_{2}$ was discovered amongst the binary
compounds- and the vast phase space of quaternary compounds has barely
been scratched by the materials physicist.  This is a frontier
of exponentiating possibilities, forming a glorious continuum 
spanning from the simplest collective
properties of the elements, out towards the most dramatic emergent
phenomenon of all- that of life itself. 

Curiously, this new frontier continues to preserve its links with technology and
applications.  During the past four decades, the size of
semi-conductor memory has halved every 18 months, following Moore's law\cite{moore}. Extrapolating this
unabated trend into the future, sometime 
around 2020, the number of atoms required to store a single bit of
information will reach unity, forcing technology into the realm of the
quantum. Just as the first industrial revolution of the early 19th century
was founded on the physical principles of thermodynamics, and the
wireless and television revolutions of the 20th century were built
largely upon the understanding of classical electromagnetism, we can
expect that technology of this new century will depend on the new
principles- of collective and quantum mechanical behavior that our
field has begun, and continues to forge today. 

\section*{Acknowledgments}

Much of the content in this article was first presented as a talk at
the 1999 centennial APS meeting in Atlanta.  
I am indebted to P. W. Anderson
for the analogy with the gold atom, and would like to thank M. E.
Fisher for pointing out that many of the key developments derive from
revisiting areas once thought to be closed. 
I have benefited from discussions and email
exchanges with many people, including E. Abrahams, P. W. Anderson,
G. Baym, P. Chandra, M. Cohen, M. E. Fisher, K. Levin, G. Lonzarich, R. Laughlin and D. Pines. This research has been been supported by
the NSF through grants DMR 9983156 and DMR 0312495.

\vspace{.5cm}
\noindent{Piers Coleman\\
Center for Materials Theory\\
Rutgers University\\
Piscataway, NJ 08855, U.S.A. }

\end{document}

%% file: table1x.tex
\begin{center}
{\bf Table. 1. Order parameters, broken symmetry  and rigidity.}
\vskip 0.2 truein
{
\begin{tabular}{|c||c|c|}
\hline
Name  & Broken Symmetry & Rigidity/Supercurrent
\\
\hline\hline
&&\\
Crystal& Translation Symmetry & Momentum superflow\\
&& (Sheer stress)
\\
\hline
&&\\
Superfluid & Gauge symmetry & Matter superflow\\
&&\\
\hline
&&\\
Superconductivity & E.M. Gauge symmetry & Charge 
superflow\\
&&\\
\hline
&&\\
Ferro and Anti-ferromagnetism & Spin rotation symmetry & Spin 
superflow\\
&&(x-y magnets only)\\
\hline
&&\\
Nematic Liquid crystals  & Rotation symmetry  & Angular momentum \\
&&superflow \\
\hline
&&\\
?  & Time Translation Symmetry  & Energy superflow ?\\
&&\\
\hline
\end{tabular}
}
\end{center}

\noindent